\documentstyle[aps,prc,epsfig]{revtex}

\begin{document}
\draft
\title{A quantum Monte-Carlo method for fermions,
       free of discretization errors.}
\author{S.M.A. Rombouts \thanks{Postdoctoral Fellow of the Fund for Scientific 
                            Research - Flanders (Belgium)},
        K. Heyde and N. Jachowicz}
\address{Universiteit Gent, Vakgroep Subatomaire en Stralingsfysica
         \\
         Proeftuinstraat 86, B-9000 Gent, Belgium
         \\
         E-mail: Stefan.Rombouts@rug.ac.be, Kris.Heyde@rug.ac.be
         }
\date{May 20, 1998}
\maketitle
\begin{abstract}
\noindent
In this work we present a novel quantum Monte-Carlo method for 
fermions, based on an exact decomposition of the Boltzmann operator
$exp(-\beta \hat{H})$.
It can be seen as a synthesis of several related methods.
It has the advantage that it is free of discretization errors,
and applicable to general interactions,
both for ground-state and finite-temperature calculations.
The decomposition is based on low-rank matrices, 
which allows faster calculations.
As an illustration, the method is applied to an analytically solvable model
(pairing in a degenerate shell) and to the Hubbard model.
\end{abstract}
\pacs{02.70.Lq,21.60.Ka,71.10.Fd}

Quantum Monte-Carlo methods offer an interesting way to obtain numerical 
results for large quantum systems \cite{ceperley1,ceperley2,lind}. 
A number of Monte-Carlo methods, that go by names as  
auxiliary-field \cite{sugiyama}, 
shell-model \cite{smmc}, 
grand-canonical \cite{gcmc} or 
projection quantum Monte-Carlo \cite{lind}
are based on the decomposition of the Boltzmann operator 
$exp(-\beta \hat{H})$ as a sum or integral over
exponentials of one-body operators. 
The latter are easy to handle numerically.
Simple algebraic expressions exist to calculate their grand-canonical 
\cite{gcmc} or canonical trace \cite{smmc,tracec}
or their overlap between Slater determinants \cite{lind,white}.
The sum or integral is then evaluated using Monte-Carlo techniques,
most often Markov-chain Monte-Carlo techniques such 
as the Metropolis algorithm \cite{metropolis}.
The basic ingredients of such a decomposition are the 
Suzuki-Trotter decomposition \cite{suzuki} 
to separate non-commuting parts of the Hamiltonian 
and the Hubbard-Stratonovich transform \cite{hubbard,straton}
to linearize the two-body part of the Hamiltonian.
Both ingredients lead to systematic discretization errors in the calculations.
Furthermore, for general two-body interactions these methods require 
many manipulations with dense matrices and hence a lot of CPU time. 

A finite-temperature method that is free of discretization errors 
was presented by Cerf \cite{cerf}. 
It is based on the expression
\begin{equation}
 e^{-\beta \left( \hat{H}_0 + \hat{V} \right)} =  
    e^{-\beta \hat{H}_0} + \;
 \sum_{m=1}^{\infty} (-\beta)^m 
 \int_{0}^{\beta} dt_m \cdots \int_{0}^{t_2} dt_1 \; 
  e^{- t_1 \hat{H}_0} \hat{V}
  e^{- (t_2-t_1) \hat{H}_0} \hat{V} \cdots
  e^{- (t_m-t_{m-1}) \hat{H}_0} \hat{V} e^{-(\beta - t_m) \hat{H}_0}.
\label{cerfexp}
\end{equation}
However, this method is limited to very specific types of interactions.
It has only been applied to a model with pairing in a degenerate shell.
Cerf's method is not based on a decomposition into a sum of exponentials 
of one-body operators.
Hence the simple algebraic expressions for the (grand-)canonical trace
and for the overlap between Slater determinants cannot be applied here.

In this letter we present a new decomposition of the Boltzmann operator
that is free of discretization errors, just like Cerf's method,
but that is applicable to general interactions and moreover results
in a (partially continuous) sum over exponentials of one-body operators,
just like auxiliary-field quantum Monte-Carlo methods.
It can be seen as a synthesis of the two.
Furthermore this method is based on low-rank matrices, 
which results in a considerable speed-up in comparison with standard
auxiliary-field quantum Monte-Carlo methods.
The decomposition is an exact one, so the only error in the calculations
is the statistical error originating from the Monte-Carlo sampling
of the terms in the decomposition
(apart from the round-off error due to the limited machine precision).
Because the method is applicable to general fermionic two-body interactions,
it can be of use for nuclear, atomic 
as well as condensed-matter physics problems. 
However, it is not free of sign problems at low temperatures.
But just like the standard Quantum Monte-Carlo methods,
there is no sign problem for the attractive Hubbard model,
the repulsive Hubbard model at half filling \cite{lind} 
and the mean-field plus pairing model for even-even atomic nuclei 
\cite{cerf,lang,pairqmc}.

The trick to arrive at the decomposition of the Boltzmann operator,
is to replace $-\beta \hat{V}$ in expression \ref{cerfexp} 
with $\mu -\beta \hat{V}$,
where $\mu$ is an arbitrary, real positive parameter.
Adding $\mu$ to the exponent has no influence on the properties
calculated with the Boltzmann operator, 
and it can simply be corrected by a factor $e^{-\mu}$.
Instead of expression \ref{cerfexp} we now obtain:
\begin{equation}
  e^{-\beta \hat{H}_0 + \left( \mu - \beta \hat{V} \right)} =
   \left[1+ \sum_{m=1}^{\infty} \mu^m 
         \int_{0}^{1} dt_m \cdots \int_{0}^{t_2} dt_1 \; 
         \prod_{i=1}^m \left(1- \frac{\beta}{\mu}
         e^{-t_i \beta \hat{H}_0} \hat{V} e^{t_i \beta \hat{H}_0} \right)
   \right]  e^{- \beta \hat{H}_0}, 
\label{myexp2}
\end{equation}
This last expression is reminiscent of the expression for the partition
function in the interaction representation derived in \cite{fetter},
with the difference that here the factors have the 
form $1-\frac{\beta}{\mu}\hat{V}(t \beta)$ instead of $-\hat{V}(t)$
(where $\hat{V}(t)$ is the two-body Hamiltonian in the interaction
 representation).
It is this extra constant $1$ that allows to make a decomposition
into a sum of exponentials of one-body operators.
To achieve this we start by constructing a decomposition for
$1- \frac{\beta}{\mu} \hat{V}$.
Hereby we build on expressions derived in reference \cite{dischubs}.

For the pairing Hamiltonian one has
\begin{equation}
 \hat{V}= - G \sum_{k,k'>0}
            \hat{a}_{k'}^{\dagger} \hat{a}_{\bar{k'}}^{\dagger}
            \hat{a}_{\bar{k}}\hat{a}_{k},
\label{pairingv}
\end{equation}
where the operator $\hat{a}_{k}^{\dagger}$ creates a particle in 
the corresponding single-particle state
and with $\bar{k}$ the time-reversed state of the state $k$.
The notation $k,k'>0$ denotes that the summation for $k$ and $k'$ 
should run over states with angular momentum projection  $j_z>0$ only.
Using lemma 1 from reference \cite{dischubs}, we obtain
\begin{equation}
1- \frac{\beta}{\mu} \hat{V} = 
 \frac{1}{2 \Omega^2}
 \sum_{k,k'>0} 
 \sum_{s=\pm 1}
\hat{\cal O} \left(1+s \gamma \, A_{k'}^{\dagger} A_{k} 
                    +s \gamma \, A_{\bar{k'}}^{\dagger} A_{\bar{k}} \right),
\end{equation}
with $\gamma= \Omega \sqrt{\frac{\beta G}{\mu}}$
and $\Omega$ half the number of single-particle states.
$A_{k}$ is the row matrix
with a one on the entry corresponding to the state $k$, 
and zero's anywhere else.
As defined in reference \cite{dischubs}, for a square matrix $Q$, 
the operator $\hat{\cal O} (Q)$ transforms a Slater determinant 
$\Psi_M$, represented by the matrix $M$, into the Slater  $\Psi_{M'}$,
with $M'=Q \, M$.
If $Q$ is non-singular, then $\hat{\cal O} (Q)$ is the exponential
of a one-body operator.
Note that this decomposition has a symmetry between the states $k>0$
and their time-reversed states. 
This symmetry prevents sign problems for even particle numbers.

For the repulsive Hubbard model one can  take
\begin{equation}
 \hat{V} = U \sum_i \left( \hat{n}_{\uparrow i} \hat{n}_{\downarrow i}
         - \frac{\hat{n}_{\uparrow i} +\hat{n}_{\downarrow i}}{2} \right).
\end{equation}
Then 
\begin{equation}
1- \frac{\beta}{\mu} \hat{V} =
\frac{1}{2 N_S} \sum_i  \left[
 e^{\gamma ( \hat{n}_{\uparrow i} - \hat{n}_{\downarrow i})}
+ e^{-\gamma ( \hat{n}_{\uparrow i} - \hat{n}_{\downarrow i})} \right],
\end{equation}
provided that $\cosh(\gamma)=1+ \frac{U \beta N_S}{2 \mu}$,
with $N_S$ the number of lattice sites 
and $\hat{n}_{\sigma i}=\hat{a}^{\dagger}_{\sigma i}\hat{a}_{\sigma i}$. 
Lemma 1 from reference \cite{dischubs} allows one to construct 
analogous {\em exact} decompositions of $1- \frac{\beta}{\mu} \hat{V}$, 
based on matrices of low rank, for any fermionic interaction.

\begin{figure}
\centerline{
\begin{picture} (230,80)
  \thinlines
  \put (30,20) {\line(1,0){170}}
  \multiput (200,20) (5,0) {4} {\line(1,0){3}}
  \put (30,60) {\line(1,0){170}}
  \multiput (200,60) (5,0) {4} {\line(1,0){3}}
  \put (10,20) {\begin{picture}(16,60)
	            \put (0,40){\makebox(16,20){$\hat{U}_{I,\tau}:$}}
	            \put (0,20){\makebox(16,20){$I_i:$}}
	            \put (0, 0){\makebox(16,20){$\tau_i:$}}
                 \end{picture}}
  \put ( 30,20) {\begin{picture}(10,40)
                    \put (0,0) {\line(0,1){40}}
	            \put (0,20){\makebox(10,20){$0$}}
	            \put (0, 0){\makebox(10,20){$\tau_{1}$}}
		 \end{picture}}
  \put ( 40,20) {\begin{picture}(15,40)
                    \put (0,0) {\line(0,1){40}}
	            \put (0,20){\makebox(15,20){$0$}}
	            \put (0, 0){\makebox(15,20){$\tau_{2}$}}
		 \end{picture}}
  \put ( 55,20) {\begin{picture}(28,40)
                    \put (0,0) {\line(0,1){40}}
	            \put (0,20){\makebox(28,20){$0$}}
	            \put (0, 0){\makebox(28,20){$\tau_{3}$}}
		 \end{picture}}
  \put ( 83,20) {\begin{picture}(19,40)
                    \linethickness{3pt}
                       \put (0,0) {\line(0,1){40}}
                    \thinlines
	            \put (0,20){\makebox(19,20){$I_4$}}
	            \put (0, 0){\makebox(19,20){$\tau_{4}$}}
		 \end{picture}}
  \put (102,20) {\begin{picture}(11,40)
                    \put (0,0) {\line(0,1){40}}
	            \put (0,20){\makebox(11,20){$0$}}
	            \put (0, 0){\makebox(11,20){$\tau_{5}$}}
		 \end{picture}}
  \put (113,20) {\begin{picture}(23,40)
                    \put (0,0) {\line(0,1){40}}
	            \put (0,20){\makebox(23,20){$0$}}
	            \put (0, 0){\makebox(23,20){$\tau_{6}$}}
		 \end{picture}}
  \put (136,20) {\begin{picture}(28,40)
                    \put (0,0) {\line(0,1){40}}
	            \put (0,20){\makebox(28,20){$0$}}
	            \put (0, 0){\makebox(28,20){$\tau_{7}$}}
		 \end{picture}}
  \put (164,20) {\begin{picture}(16,40)
                    \linethickness{3pt}
                       \put (0,0) {\line(0,1){40}}
                    \thinlines
   	            \put (0,20){\makebox(16,20){$I_8$}}
	            \put (0, 0){\makebox(16,20){$\tau_{8}$}}
		 \end{picture}}
  \put (180,20) {\begin{picture}(16,40)
                    \put (0,0) {\line(0,1){40}}
	            \put (0,20){\makebox(16,20){$0$}}
	            \put (0, 0){\makebox(16,20){$\tau_{9}$}}
		 \end{picture}}
  \put (196,20) {\begin{picture}(20,40)
                    \put (0,0) {\line(0,1){40}}
	            \put (0,20){\makebox(20,20){$\cdots$}}
	            \put (0, 0){\makebox(20,20){$\cdots$}}
                 \end{picture}}
  \put (83,6) {\begin{picture}(81,12)
                    \put (21,6) {\vector(-1,0){21}}
                    \put (21,4) {\line(0,1){4}}
                    \put (60,6) {\vector( 1,0){21}}
                    \put (60,4) {\line(0,1){4}}
                    \put (40,6) {\makebox(0,0) {$t_2-t_1$}}
	       \end{picture}}
  \put ( 83,60) {\begin{picture}(0,20)
                    \put(0,16){\makebox(0,0){ $Q_{I_4}$}}
                    \put(0,10){\vector(0,-1){8}}
                 \end{picture}}
  \put (164,60) {\begin{picture}(0,20)
                    \put(0,16){\makebox(0,0){ $Q_{I_8}$}}
                    \put(0,10){\vector(0,-1){8}}
                 \end{picture}}
  \put ( 30,62) {\makebox(53,12){\small $e^{- t_1\beta \hat{H}_0}$}}
  \put ( 83,62) {\makebox(81,12){\small $e^{- (t_2-t_1)\beta \hat{H}_0}$}}
  \put (174,62) {\makebox(52,12){\small $e^{- (t_3-t_2)\beta \hat{H}_0}$}}
\end{picture}}
\caption{A schematic way to represent the terms in the decomposition
         \ref{myexp2}. 
         The array $\tau_i$ represents the inverse-temperature intervals,
         the array $I_i$ represents the parts of $1-\frac{\beta}{\mu}\hat{V}$
         that are inserted.}
\label{figure1}
\end{figure}
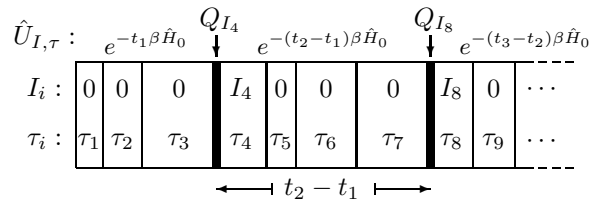

The Monte-Carlo algorithm has to sample over
all values of $m$ between 0 and infinity,
over all possible sets  $0 \leq t_1 \leq \cdots \leq t_m \leq 1$
and at each interval over all the terms in the decomposition of
$1- \frac{\beta}{\mu} \hat{V}$.
To perform this sampling numerically, a large number of intervals is taken. 
Let $N_x$ be the number of intervals.
To each interval $i$ we assign a fraction $\tau_i$ of the inverse temperature,
such that $\sum_{i=1}^{N_x} \tau_i = 1$,
and an index $I_i$ to indicate that
part of the decomposition of $1-\frac{\beta}{\mu}\hat{V}$ 
that is inserted at that interval.
If no part is inserted, then $I_i=0$.
In total there are $m$ out of $N_x$ intervals with $I_i \neq 0$.
Thus a term of order $m$ in the decomposition \ref{myexp2} is
represented by a configuration with $m$ intervals for which $I_i \neq 0$.
The sum of the coefficients $\tau_i$ between 
the $j^{th}$ and the $(j+1)^{th}$ insertion of $1-\frac{\beta}{\mu}\hat{V}$,
has to be equal to $t_{j+1}-t_j$.
This scheme is visualized in figure \ref{figure1}.

This representation is not unique. 
To obtain  every combination with the right weight, we have to take
into account an extra weight factor $(N_x-m)!/N_x!$ 
for a configuration of order $m$.
The operator corresponding to a particular configuration 
can then be calculated as
\begin{equation}
 \hat{U}_{I,\tau}= \prod_{i=1}^{N_x} \left( e^{-\tau_i \beta \hat{H}_0} 
 \hat{Q}_{I_i} \right),
\end{equation}
with $\hat{Q}_j$ the $j^{th}$ part in the decomposition of 
$1-\frac{\beta}{\mu} \hat{V}$, and $\hat{Q}_0=1$.
From a computational point of view it is advantageous to work in the
interaction representation, i.e.
\begin{equation}
 \hat{U}_{I,\tau}= \left( \prod_{i=1}^{N_x} \hat{Q}_{I_i}(t_i) \right) 
                   e^{- \beta \hat{H}_0} ,
\end{equation}
with $t_i = \sum_{k=1}^{i} \tau_k$ 
and  $\hat{Q}_j(t)$ the $j^{th}$ part in the decomposition of 
$1-\frac{\beta}{\mu} \hat{V} (t \beta)$.
This decomposition is obtained by multiplying 
the row matrices used in the decomposition of $1- \frac{\beta}{\mu} \hat{V}$,
with the matrix $e^{\pm t \beta H_0}$,
where $H_0$ is the matrix representation of the one-body operator
$\hat{H_0}$ in the single-particle space.
The operator $\hat{U}_{I,\tau}$ is a product of exponentials of
one-body operators and thus the exponential of a one-body operator itself.
Therefore, one can easily calculate its grand-canonical \cite{gcmc}
or canonical trace \cite{tracec},
or apply it to a Slater determinant.
In this way, exact variants of the grand-canonical \cite{gcmc}, 
shell model \cite{smmc} and projector quantum Monte-Carlo method 
\cite{lind,sugiyama} are obtained.
When applying the grand-canonical or projector variant,
fast updating techniques analogous to the ones presented by White {\it et al.}
\cite{white} can be applied.
The rank-two structure in the decomposition of the factors
$1- \frac{\beta}{\mu} \hat{V} (t \beta)$,
allows to make a quick update requiring only 
$8 N_s^2$ flops, even for general interactions.
These updates have to be performed only when $I_{i} \neq 0$,
On average, this amounts to $\langle m \rangle_{MC}$ times 
to update the whole configuration,
where $\langle m \rangle_{MC}$ is value of the order $m$ in the decomposition
\ref{myexp2}, averaged over all Monte-Carlo samples.
For the canonical algorithm such a fast updating technique is not possible.
There the performance can be drastically improved using 
{\it guided sampling} \cite{thesis}.
Instead of the canonical ($N$-particle) weight 
$w_{I,\tau}=\mu^m (N_x-m)!
            \mbox{Tr}_{N} \left(\hat{U}_{I,\tau} \right) / N_x!$, 
one uses a local approximation $\tilde{w}_{I,\tau}$ that allows fast updates.
After a number of steps these updates are then accepted or rejected
collectively, 
according to the ratio $q=\frac{\tilde{w}_{I,\tau}}{\tilde{w}'_{I,\tau}}
                        \frac{w'_{I,\tau}}{w_{I,\tau}}$.
Using a generalized Metropolis algorithm \cite{hastings},
that includes the factor $\mu^m \frac{(N_x-m)!}{N_x !}$ in the
proposition probability, one can set up a very efficient Markov chain,
with acceptance rates close to unity and with autocorrelation lengths 
of a few sweeps.

Because the updating procedure is the most time-consuming step 
of the algorithm, the required CPU-time will be proportional to
$\langle m \rangle_{MC}$. 
Therefore it is important to have a good estimate of this quantity in advance.
If the grand-canonical or canonical algorithm is used,
and if the weight $w_{I,\tau}$ is positive for all configurations,
then one can show that 
\begin{equation}
\langle m \rangle_{MC} 
 = \mu - \beta \langle \hat{V} \rangle_{\beta}.
\label{muv}
\end{equation}
This shows that the CPU-time is proportional to the parameter $\mu$,
but also to the thermal expectation value 
$\langle \hat{V} \rangle_{\beta}$ of the residual interaction.
Though $\mu$ is an arbitrary parameter, we have experienced
that a good balance between low CPU-cost per sweep 
and fast mixing of the Markov chain is obtained by taking 
$\mu \simeq \beta | \langle \hat{V} \rangle_{\beta}|$.
Note that one has to take $N_x$ such that it is always larger than 
the largest value of $m$ encounterd during the Monte-Carlo sampling.
Apart from that, the method and the CPU-time it requires 
is independent of $N_x$.
Expression \ref{muv} can also be used to obtain a value for 
$\langle \hat{V} \rangle_{\beta}$ from the Monte-Carlo calculation.
If $w_{I,\tau}$ can become negative, then one has to take into account
the sign of $w_{I,\tau}$ in expression \ref{muv}.
To calculate expectation values for other observables, 
one has to use the techniques developed for auxiliary-field quantum 
Monte-Carlo methods, as described in references \cite{smmc,white,pairqmc}.

\begin{figure}
\epsfig{file=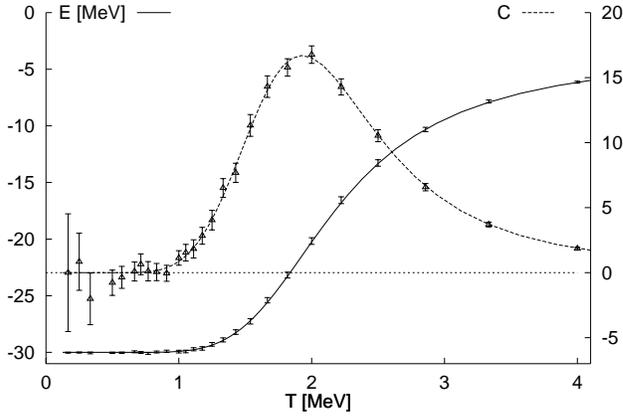, width=9cm}
\caption{Energy $E$ and specific heat $C$ 
         as a function of temperature $T$,
         for a model with pairing in a degenerate shell,
         as described in the text.
         The curves correspond to the analytical results,
         the errorbars indicate $2 \times 2 \sigma$ 
         intervals for the Monte-Carlo results.}
\label{figure2}
\end{figure}

To test our quantum Monte-Carlo method, we have applied it to a
model with pairing in a degenerate shell.
This many-body problem can be solved analytically using the 
seniority scheme \cite{ring}.
We considered a system with 10 particles
in a degenerate shell ($\hat{H}_0=0$) of 20 single-particle states.
We took $\hat{V}$ as in expression \ref{pairingv}, with $G=1 MeV$. 
Figure \ref{figure2} shows the results for the energy 
and the specific heat of the model, in the canonical ensemble.
They agree perfectly with the analytical results.
These observables were evaluated using
equations 11 and 14 from reference \cite{pairqmc}.

\begin{figure}
\epsfig{file=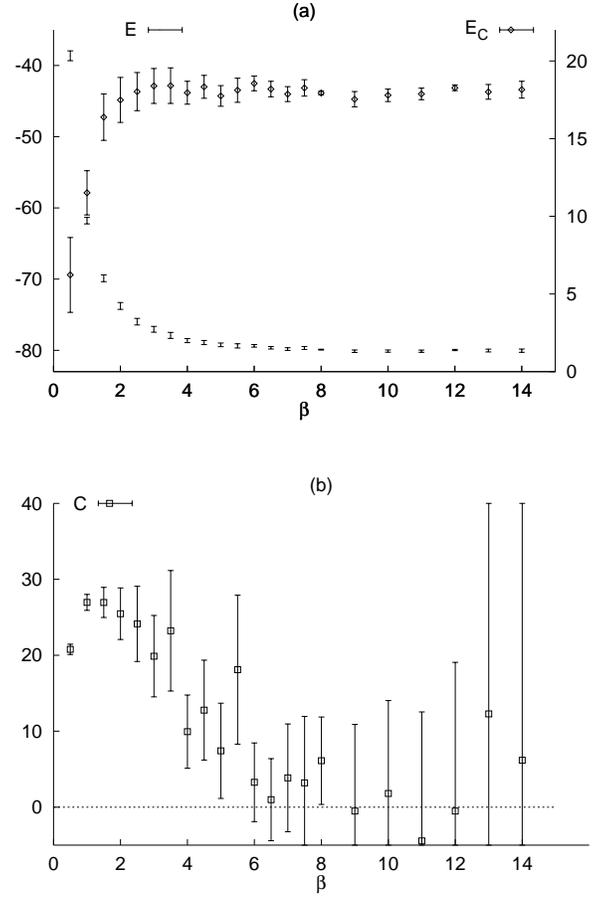, width=9cm}
\caption{\, Total energy $E$ and Coulomb energy
         $E_C=\langle \hat{V} \rangle_{\beta}+ U N/2$,  (a),
         and specific heat $C$, (b),
         as a function of inverse temperature $\beta$,
         for the Hubbard model described in the text,
         with 28 spin-up and 28 spin-down particles 
         on an 8-by-8 lattice, $U=2|t|$.
         Units were chosen such that $t=1$.
         ($2\sigma$ errorbars)}
\label{figure3}
\end{figure}

\begin{figure}
\epsfig{file=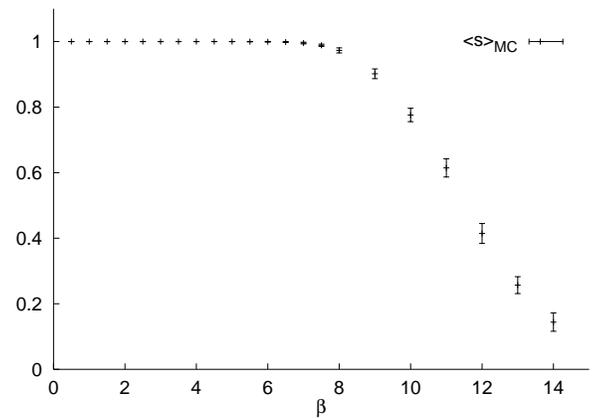, width=9cm}
\caption{Average sign $\langle s \rangle_{MC}$
         as a function of inverse temperature $\beta$,
         for the same model as figure \ref{figure3}.
	 ($2\sigma$ errorbars)}
\label{figure4}
\end{figure}

To demonstrate the versatility of the method, we have also applied it 
to the one-band repulsive Hubbard model.
We have calculated the energy, specific heat and Coulomb energy 
in the canonical ensemble with 28 spin-up and 28 spin-down particles,
for a system with nearest-neighbour hopping only (strength $t=1$), 
on an 8-by-8 lattice with periodic boundary conditions 
and an interaction strength $U=2|t|$.
The results are shown in figure \ref{figure3}.
The average sign of the weight $w_{I,\tau}$ for this system
is shown as a function of the inverse temperature $\beta$ in figure 
\ref{figure4}.
All calculations were done on a Pentium Pro 200 MHz PC 
and a Digital Alphastation 255/300MHz workstation.

Just like other quantum Monte-Carlo methods for fermions,
our method is not free of sign problems at low temperatures.
However, for a number of systems, a symmetry exists 
that guarantees good sign properties. 
This is among others the case for the attractive Hubbard model,
the repulsive Hubbard model at half filling
and the mean-field plus pairing model for even-even atomic nuclei.
For other systems, one can expect the sign properties to be different
from standard quantum Monte-Carlo methods,
because a different decomposition for the Boltzmann operator is used.
This might or might not be an improvement, 
depending on the specific situation,
but has to be studied more deeply.

The method can be improved by performing a (finite temperature) Hartree-Fock
calculation for the Hamiltonian $\hat{H}$ in advance.
This results in a one-body part $\hat{H}_0$ and a residual 
interaction $\hat{V}$ such that $|\langle \hat{V} \rangle_{\beta}|$ is reduced.
From expression \ref{muv} it is clear that this reduces 
$\langle m \rangle_{MC}$ and hence reduces the CPU-time.
Furthermore, one can expect that this will improve the average sign. 

In conclusion, we have developed a quantum Monte-Carlo method for
fermions based on an exact decomposition of the Boltzmann operator 
into a continuous sum over exponentials of one-body operators. 
Because of the low-rank matrices used in the decomposition,
fast matrix multiplications and efficient updating procedures can be applied.
The method can be applied to systems with general two-body interactions,
in the grand-canonical or canonical ensemble, or with ground-state projection.
It allows to calculate thermal and ground-state expectation values
for any observable.
It is an exact method, apart from statistical errors.

The authors are grateful to the F.W.O. (Fund for Scientific Research) 
- Flanders and to the Research Board of the University of Gent for
financial support.

\twocolumn
\vspace{3cm}

\end{document}